\documentclass[a4paper]{spie}  %>>> use this instead for A4 paper
\usepackage[]{graphicx}

\title{Gamma--ray lens development status for a European
Gamma--Ray Imager} 

\author{F. Frontera\supit{a,d}, A. Pisa\supit{a}, V. Carassiti\supit{b},
F. Evangelisti\supit{b}, G. Loffredo\supit{a}, D. Pellicciotta\supit{a},
 K.H. Andersen\supit{c}, P. Courtois\supit{c}, L. Amati\supit{d}, 
E. Caroli\supit{d}, T. Franceschini\supit{d},
G. Landini\supit{d}, S. Silvestri\supit{d}, J.B. Stephen\supit{d}
\skiplinehalf
\supit{a}University of Ferrara, Physics Department, Via Saragat 1, 44100
Ferrara, Italy; \\
\supit{b}Istituto Nazionale Fisica Nucleare, Sezione di Ferrara, Via Saragat 1,
44100 Ferrara, Italy; \\
\supit{c}Institute Laue--Langevin, 6 Rue Jules Horowitz, 38042 Grenoble, France
\supit{d}INAF, IASF Bologna, Via Gobetti 101, 40129 Bologna, Italy
}

\authorinfo{Further author information: (Send correspondence 
to F.F.)\\ F.F: E-mail: frontera@fe.infn.it, Telephone: +39 0532 974 254}

\begin{document} 
\maketitle 

\begin{abstract}
A breakthrough in the sensitivity level of the hard X-/gamma-ray 
telescopes, which today are based on detectors that view the sky through 
(or not) coded masks, is expected when focusing optics will be 
available also in this energy range. Focusing techniques are now 
in an advanced stage of development. To date the most efficient 
technique to focus hard X-rays with energies above 100 keV appears 
to be the Bragg diffraction from crystals in transmission 
configuration (Laue lenses). Crystals with mosaic structure appear 
to be the most suitable to build a Laue lens with a broad 
passband, even though other alternative structures are being 
investigated. The goal of our project is the development of a 
broad band focusing telescope based on gamma-ray lenses for 
the study of the continuum emission of celestial sources from 
60 keV up to $>$600 keV. We will report details of our project, 
its development status and results of our assessment study 
of a lens configuration for the European Gamma Ray Imager (GRI) 
mission now under study for the ESA plan "Cosmic Vision 
2015-2025".

\end{abstract}

\keywords{Laue lenses, gamma-ray instrumentation, focusing
telescopes, gamma-ray observations}

\section{INTRODUCTION}
\label{sect:intro}  

As demonstrated by the long history of astronomy,
any progress in the knowledge of the sky is strictly correlated with 
an increase in the instrument sensitivity. In hard ($>$10 keV) 
X--ray astronomy, focusing telescopes now under development are
expected to provide a big leap in our knowledge of the high energy 
astrophysical phenomena. While below 100 keV X-ray optics based 
on multilayer coatings (ML) appears to be the best technique to build hard
X--ray mirrors,
above 100 keV the best candidate technique at the present time appears the Bragg 
diffraction from crystals in transmission configuration (Laue lenses). 

A huge increase in sensitivity (by a factor 10--100)
above 100 keV would open a new window of investigation, with the concrete possibility
of fixing many open issues, apart from the prospect of unexpected discoveries.
Among the open issues which can be fixed only with high energy ($>100$~keV) 
observations, we mention (see  Frontera et al.\cite{Frontera05}):
\begin{itemize}
\item
Origin and physics of the high energy cut-offs of type 1 and type 2 
Active Galactic Nuclei (AGNs);
\item
Contribution of (thermal and non-thermal) AGNs to the high energy Cosmic X-/gamma-ray 
background (CXB). Synthesis models 
require a spectral roll-over with an e-folding energy of 100-400 keV in AGNs.  So far 
only a few sparse measurements are available.
\item
Role of the antimatter in the Universe from the study of the $e^+-e^-$ 
pair annihilation line and the effect on it of strong gravitational fields;
\item
Role of non thermal mechanisms in extended objects (Supernova remnants, 
Galaxy clusters);
\item
Properties of the high energy emission in presence of super-strong 
($>10^{13}$ G) magnetic fields (mass accreting X-ray pulsars, anomalous X-ray 
pulsars, Soft Gamma-Ray Repeaters);
\item
Role of Comptonization in compact mass accreting objects and 
in Gamma Ray Bursts;
\item
Study the star explosion mechanisms via the detection  of the nuclear 
lines from synthesized elements (e.g., Co, Ni, Ti);
\end{itemize}

Here we give the status of our HAXTEL (= HArd X-ray TELescope) 
project devoted to  develop Laue lenses for the study of the continuum 
spectra from 60 to 600 keV. This development is complementary to that 
which is being performed by other groups\cite{Vonballmoos04}, 
devoted to the development of Laue lenses for the study of nuclear 
gamma--ray lines. All these studies will be usefully exploited for the study
of a Gamma Ray Imager (GRI) mission now under study to be submitted to the 
European Space Agency (ESA) for its mission plan ''Cosmic Vision 2015-2025''.

%%%%%%%%%%%%%%%%%%%%%%%%%%%%%%%%%%%%%%%%%%%%%%%%%%%%%%%%%%%%%

\section{Results obtained so far}
Results of the activity perfomed thus far are the subject of various
papers \cite{Pisa04,Pisa05,Pellicciotta06}. They concern a theoretical
feasibility study devoted to establish the best design of a Laue lens telescope 
along with its sensitivity expectations \cite{Pisa04,Pisa05}, Monte Carlo
simulations of the expected optical properties of Laue lenses, 
reflectivity measurements of mosaic crystal samples of Cu(111) \cite{Pellicciotta06}.
Here we summarize the most relevant results.
 
\subsection{Results from a feasibility study}
\label{study}

This study was performed through calculations.
The Laue lens of our project is requested to have  a spherical shape 
with radius $R$ and focal length $f = R/2$. Assuming flat crystals, in order to 
get the best approximation of the lens geometry, crystal 
tiles with small facets are required (of the order of $10\times 10$~mm$^2$). 

Perfect crystals are not suitable for building lenses for astrophysical 
investigations. Indeed they show a high reflection efficiency but in 
a very narrow energy band (of the order of a few eV).  Given that we have to cover
with good  reflection efficiency a relatively broad 
energy  band (several hundreds of keV), special crystals, with properly controlled 
lattice deformations, appear to be more useful. Crystals of this kind
include mosaic crystals, bent crystals and crystals with non constant lattice 
spacing $d$ induced by doping materials or thermal gradients.
For our project we have assumed mosaic crystals, made of crystallites misaligned
each with other with controlled angular spread $\beta$ (FWHM of the Gaussian-like
angular distribution of the crystallite misalignments).  The growing technique of mosaic
crystals with the desired spread is now being consolidated (e.g., Courtois et al.\cite{Courtois04}). 
In the case of a mosaic crystal, when a 
polychromatic parallel beam of X--rays  impinges on it with mean Bragg 
angle $\theta_B$, from the Bragg law ($2d\sin\theta_B=n\frac{hc}{E}$, where 
the Bragg angle $\theta_B$ is the angle between the lattice planes and the direction 
of the incident and diffracted photons, $n = 1,2, ..$ is the diffraction order, 
$E$ in keV units is the photon energy and $hc=12.4$~keV$\cdot$\AA\ if $d$ is given in \AA), 
photons in a bandwidth
%
% Equation 1
%
\begin{equation}
	\Delta E = E \, \beta / \tan \theta_B
\label{e:DeltaE}
\end{equation}
are reflected by the mosaic crystal.

Crystal tiles of thickness $t$ are assumed to have their mean lattice plane normal to 
the tile main facets, which are assumed to be square of side $l$. 

\begin{figure}[!t]
\begin{center}
\includegraphics[width=0.4\textwidth]{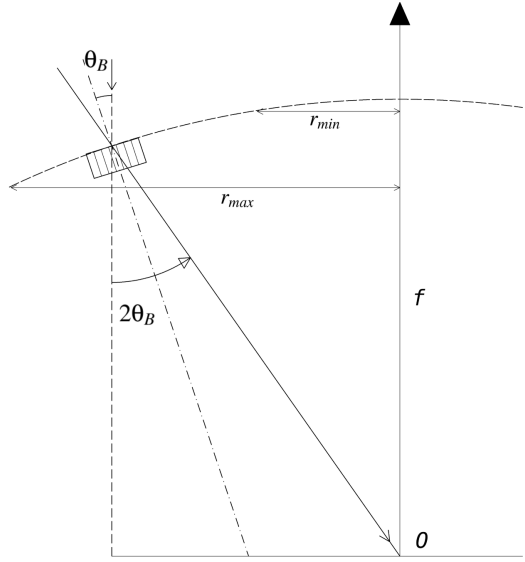}
\caption{Scheme of the Bragg diffraction in a Laue lens. The X--ray photons 
in a given energy band $\Delta E$ around $E$, which impinge on the lens 
parallel to the $z$ axis (lens axis), are diffracted only by those mosaic
crystals oriented in such a way as to satisfy Eq.~\ref{e:DeltaE}. The photons 
with centroid energy $E$ which hit the center of these crystals
are focused in the lens focus (point $O$), $r_{max}$ and $r_{min}$
define the innermost and outermost radius of the lens surface, respectively. Reprinted
from \cite{Pellicciotta06}.}
\label{f:scheme}
\end{center}
\end{figure}

To correctly focus photons (see Fig.~\ref{f:scheme}), the direction of the 
vector perpendicular to the mean lattice plane of each crystal has to intersect 
the lens axis and its inclination with respect to the focal plane has to be equal 
to the Bragg angle $\theta_B$.
The angle $\theta_B$ depends on the distance $r$ of the tile center from the lens axis 
and on the focal length $f$. For a correct focusing, it is needed that
$\theta_B = 1/2 \arctan{(r/ f)}$.
%
% Figure 1
%
Once the crystal material is established, the Bragg angle increases with $r/f$, while
the energy of the focused photons decreases with $r/f$.
More generally, once the focal length is established, the outer  and inner
lens radii, $r_{max}$ and $r_{min}$ (see Fig.~\ref{f:scheme}), depend on the
nominal energy passband of the lens ($E_{min}$, $E_{max}$) according to the 
following equation, derived from the Bragg law:  
%
% Equation 2
%
\begin{eqnarray}
	r_{max} \approx 12.4 \frac{f_{100}}{d_{hkl}({\rm \AA})\,E_{100}}\, m
\nonumber \\
	r_{min}  \approx 1.24 \frac{f_{100}}{d_{hkl}({\rm \AA})\,E_{1000}}\, m
\label{e:radii}
\end{eqnarray}
where $f_{100}$ is the focal length in units of 100 m, $d_{hkl}({\rm \AA})$ is the
spacing of the lattice planes with Miller indices (h,k,l) in units of ${\rm \AA}$, 
and $E_{100}$ and $E_{1000}$ is the photon energy in units of 100 keV and 1000 keV, 
respectively.
  
From the previous equation, it is possible to see that the lens radii not only
depend on the energy passband, but also on the crystal lattice spacing: higher
$d_{hkl}$ implies lower radii.

\begin{figure}
\begin{center}
\includegraphics[angle=-90,width=0.6\textwidth]{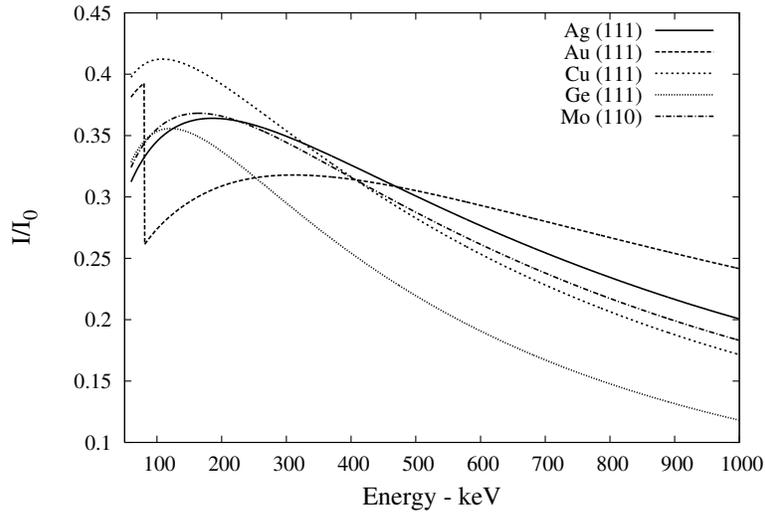}
\caption{Peak reflectivity of 5 candidate crystal materials. The Miller indices 
used give the highest reflectivity. A 
mosaic spread of 40 arcsec is assumed. The production technology of
mosaic crystals with the required spread has currently been developed only for Ge and Cu.}
\label{f:materials}
\end{center}  
\end{figure}

Among the candidate materials for their high reflectivity, Cu (111)
appears very promising for the hard X-/gamma--ray range (see Fig.~\ref{f:materials}).
It is indeed one of the 
few materials for which the technology for its growing with mosaic structure has
already been developed with good results (see below).
 
%
% Figure 2
%
For a fixed inner and outer radius, the lens passband and its effective area can
be further controled by  the use of a combination of different crystal materials.
Mosaic spread and, for a fixed material, crystal thickness are the most crucial parameters
to be fixed to optimize the lens performance. 

The effect of the crystal thickness on the lens effective area has been discussed
in previous papers (see, e.g., Pisa et al.\cite{Pisa04}). It results that 
a single thickness is not the best solution for optimizing the lens effective area in its entire 
passband.
However the optimization of the lens effective area at the highest energies could imply 
large thicknesses, that could be incompatible with lens weight constraints. 
We have investigated this issue
elsewhere \cite{Pisa06}, finding that a decrease in the thickness  of the order of 25\% with respect 
to the optimized value results in a very small decrease of the effective area, 
while the lens weight significantly decreases.
  
%
% Figure 3
%
\begin{figure}
\begin{center}
\includegraphics[angle=-90,width=0.6\textwidth]{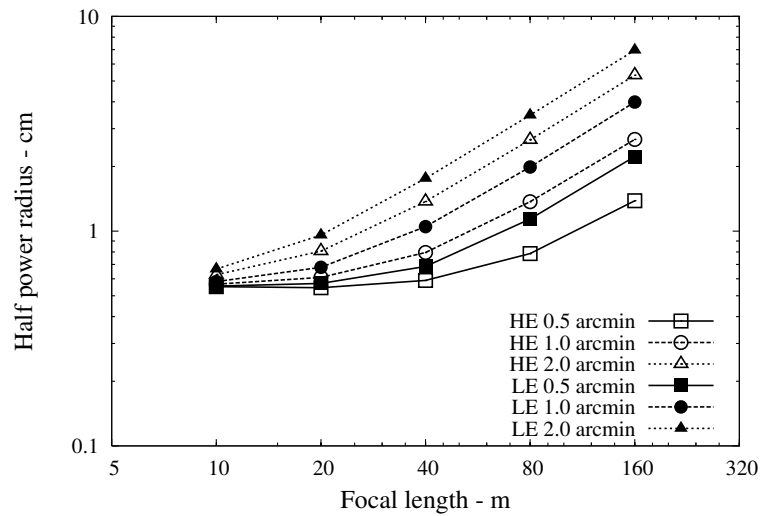}
\caption{Radius of the circle in which are focused 50\% of the reflected photons
(Half Power Radius, HPR) as a function of the focal distance, for various values of 
the mosaic spread in two energy bands, 90--110 keV (LE), and 450--550 keV (HE).} 
\label{f:hpr}
\end{center}  
\end{figure}

Also the mosaic spread $\beta$ (see Eq.~\ref{e:DeltaE}) affects the lens
performance. A higher spread gives a larger effective area \cite{Pisa04}, 
but also produces a larger defocusing of the reflected photons in the focal 
plane, as can be seen in Fig.~\ref{f:hpr}, in which we show how the radius
of the circle which contains 50\% of  the reflected photons in the focal plane (Half
Power Radius, HPR) changes with the focal length for various values of the mosaic spread.

 The best parameter which takes into account
both the effective area and the size of the focal spot in the focal plane
is the focusing factor $G$, which is given by:
\begin{equation}
G = f_{ph} \frac{A_{eff}}{A_d}
\end{equation}
 in which $A_{eff}$ is the effective area of the lens and $A_d$ is the
area of the focal spot which contains a fraction $f_{ph}$ of photons reflected by the lens. 
From the expression of the lens sensitivity to the continuum
emission
\begin{equation}
I_{min} (E) = \frac{n_\sigma}{\eta_d A_{eff} f_{ph}} \sqrt{\frac{2 B A_d}{T \, \Delta E}}
\label{e:sens}
\end{equation}
where $I_{min}$ (photons~cm$^{-2}$~s$^{-1}$~keV$^{-1}$) is the the minimum detectable 
intensity  in the interval $\Delta E$  around $E$, $n_\sigma$ is the significance
level of the signal (typically $n_\sigma = $3--5), $B$ is the focal plane detector
background intensity (counts~cm$^{-2}$~s$^{-1}$~keV$^{-1}$), $T$ is exposure time to 
a celestial source, and $\eta_d$ is the focal plane detector efficiency 
at energy $E$, it can be seen that the lens sensitivity is inversely proportional 
to $G$.

%
% Figure 4
%
\begin{figure}
\begin{center}
\includegraphics[angle=270,width=0.49\textwidth]{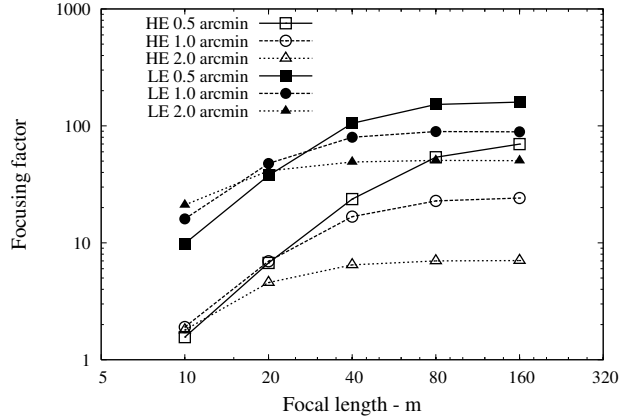}
\caption{Dependence of the focusing factor $G$ on the focal length $F$
for three different values of the mosaic spread in two different energy 
bands: 90--110 keV (LE) and 450--550 keV (HE). See text.}
\label{f:G_vs_F}
\end{center}
\end{figure}
We have investigated \cite{Frontera06} the dependence of $G$ on the focal length
for three values of the mosaic spread (0.5, 1 and 2 arcmin) and for
two different lens passbands: 90--110 keV (Low Energy, LE) and 450--550 keV (High Energy,
HE). For $f_{ph} = 0.50$ corresponding to the HPR and for crystal tiles 
of $15\times 15$~mm$^2$, we find the result shown in Fig.~\ref{f:G_vs_F}. As can be seen,  
$G$ is significantly higher 
at low energies, shows a saturation for high focal lengths,  and, 
specially at high energies, significantly increases with the spread decrease.
However, compatibly with an acceptable $G$, a larger spread would be preferred, since
it gives a higher effective area and thus a higher photon collection. In addition, a 
larger spread 
requires a lower accuracy in the positioning of the crystals in the lens \cite{Pisa04}.

Another issue we have investigated is the disposition of the mosaic crystal tiles
in the lens for a uniform coverage of the lens passband. The result is that
the best crystal tile disposition is that of an Archimedes' spiral.
This disposition \cite{Pisa04,Pellicciotta06} is crucial to get a
smooth change of  the lens effective area $A_{eff}$ with energy, apart from the
jumps due to diffraction orders higher than one. However the Archimedes' spiral
becomes less important for high focal lengths ($>$ 30 m).

The advantage of the Laue lenses with respect to the direct--viewing telescopes, is 
better illustrated by the equivalent effective area $S$, defined, in the case of a 
lens, as $S = G^2 A_d$, while, in the case of a direct--viewing telescope, $S$ 
is the useful detector area. In both cases the sensitivity (see eq.~\ref{e:sens}) 
is inversely proportional to $S^{0.5}$, but for the lenses it is proportional 
to $G$ and thus increases with the focal length (see Fig.~\ref{f:G_vs_F}). 

Special care has to be taken in the accuracy of the crystal tile positioning in the lens. 
A deviation
of the crystal orientation with respect to the nominal position degrades  the
focusing factor. The accuracy required in the crystal tile positioning to get
a negligible degradation not only
depends on the mosaic spread, as above mentioned, but also on the focal length.
Higher focal lengths require higher positioning accuracies, which is at the
current stage of development one of the major problems to be faced for
the realization of a Laue lens.

\subsection{Results from Monte Carlo simulations}

A Monte Carlo (MC) code has been developed by us to derive the  
properties of different lens configurations for either 
on--axis and off--axis incident photons. 
For a ring of square crystal tiles of $10 \times 10$~mm$^2$ cross 
section and 2 mm thickness, properly oriented for 150 keV photons, the MC results have 
already been reported \cite{Pisa06}. They confirm 
the results obtained from the theoretical investigation and extend them. 
Among the results obtained (some of them already mentioned above, see Figs.~\ref{f:hpr} 
and \ref{f:G_vs_F}), we
discuss here those concerning the optical properties of the Laue lenses, 
in particular the on-axis and off-axis responses, the lens angular resolution and its 
field of view (FOV).

\begin{figure}
\begin{center}
\includegraphics[angle=-90, width=0.8\textwidth]{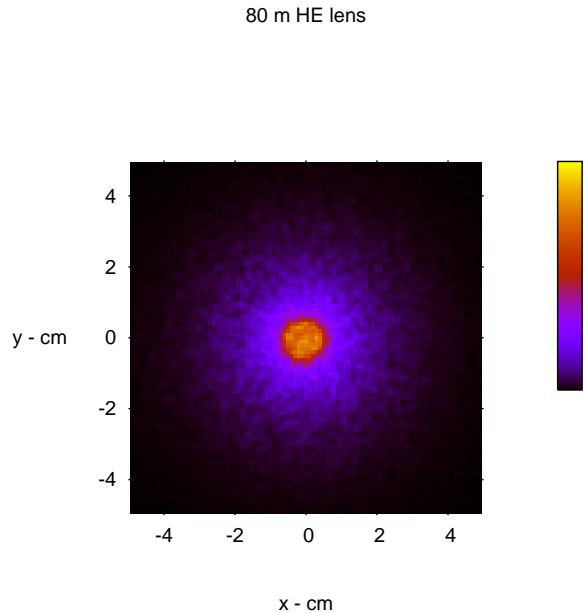}
\end{center}
\vspace{-0.5cm}
\caption{Point spread function of a lens with 80 m focal length for an on-axis source in the
 150--600 keV energy band.}
\label{f:psf_onaxis}
\end{figure}

\begin{figure}
\begin{center}
\includegraphics[width=0.5\textwidth]{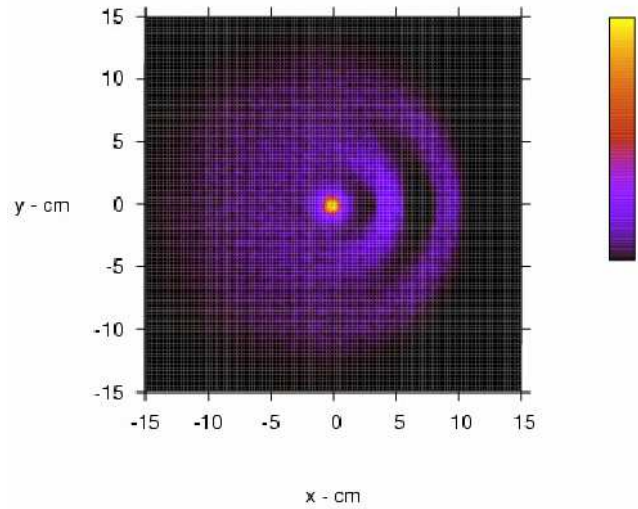}
\end{center}
\caption{Point spread function of a lens with 80 m focal length in the case of the observation
of three sources, one on-axis and the other two offset by 2 and 4 arcmin. The energy band is
150--600 keV.}
\label{f:psf_offset}
\end{figure}

\begin{figure}
\begin{center}
\includegraphics[angle=-90,width=0.5\textwidth]{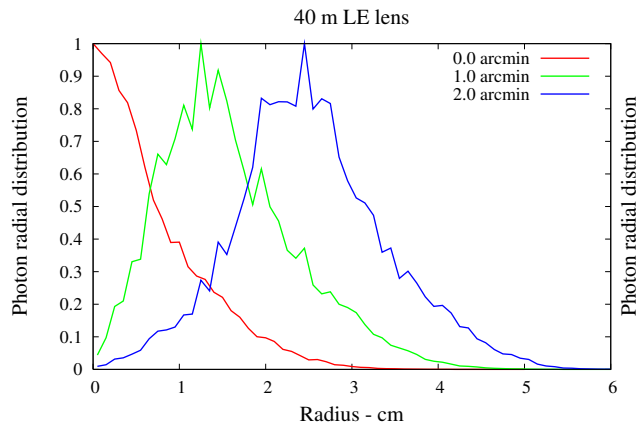}
\end{center}
\caption{Radial distribution of the photons at different offset angles for 40 m 
focal length.}
\label{f:ang}
\end{figure}

We show in Fig.~\ref{f:psf_onaxis} the expected Point Spread Function 
(PSF) of the lens with 80 m focal length in the case of an on--axis X--ray source, 
while in Fig.~\ref{f:psf_offset} we show the expected PSF when three sources are
in the lens Field of View (FOV), one on--axis and two off--axis. As can be seen, in the 
case of off--axis sources the PSF has a ring shape with
center in the lens focus, an inner radius which increases  with the offset angle
and a disuniform width with azimuth. This disuniformity
gives information on the sky direction of the X--ray source.
From the radial distribution of the focused photons due to sources with different
offset (see Fig.~\ref{f:ang}), it is possible to derive the angular resolution 
of the lens.  Conservatively, it results to be about 2 arcmin, i.e., about two times larger 
than that of the crystal mosaic spread assumed (1 arcmin). 
%
% Figure 5 
%

%
% Figure 6
%

From the numerical simulations it results that the number of photons focused by the 
lens does not significantly vary from an on--axis source and to an off-axis source. 
Thus the FOV of the lens in principle seems to be determined by the detector radius. However,
given that focused photons from sources at increasing offset spread over an increasing area,
the detector radius is determined by the lens sensitivity.

%
% Figure 7
%

\subsection{Results from experimental activity}

To test the reliability of our feasibility study, we have performed
many reflectivity measurements of mosaic crystal samples of Cu(111).
The samples, with various thickesses and mosaic spreads,  were provided by 
the Institute Laue Langevin (ILL), that has developed the technology for growing 
large single crystals of Copper with a small mosaic spread \cite{Courtois04}. The crystal samples were 
tested at the Ferrara X--ray facility \cite{Loffredo03,Loffredo04}. 
Using a pencil beam of polychromatic X--rays, reflectivity curves were derived 
in an array  of positions of the crystal cross section to study the uniformity of 
the mosaic properties. Results of these measurements have  been reported and
discussed by Pellicciotta et al. \cite{Pellicciotta06}. 
%
% Figure 8
%
\begin{figure}
\begin{center}
\includegraphics[angle=0,width=0.5\textwidth]{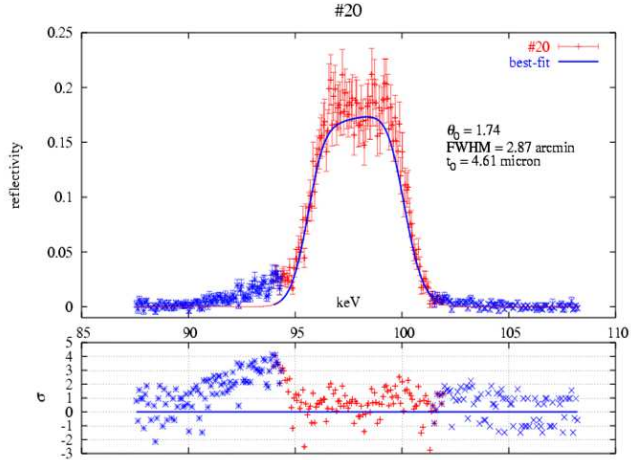}
\end{center}
\caption{Example of measured reflectivity profile.
Also shown is the best fit model with the model function derived
by \cite{Zac45} and the residuals to the model (bottom panel). 
Reprinted from Pellicciotta et al.\cite{Pellicciotta06}}
\label{f:exp}
\end{figure}

The reflectivity curves were fit with the theoretical model function derived
by Zachariasen \cite{Zac45}. In most cases the results were found satisfactory.
Figure~\ref{f:exp} shows a typical result of the fitting analysis.
Evidence of crystal inhomogeneity was observed  near the 
crystal boundaries, mainly due to the crystal cutting procedure which perturbs 
the mosaic structure.
Given that, for our goals, the crystals should be as homogeneous as possible,
the cutting procedure requires further refinement.
In spite of that, the measurements results obtained substantially
confirm the expectations of our feasibility study. 

%
% Figure 9
%
\begin{figure}
\begin{center}
\includegraphics[angle=-90,width=0.6\textwidth]{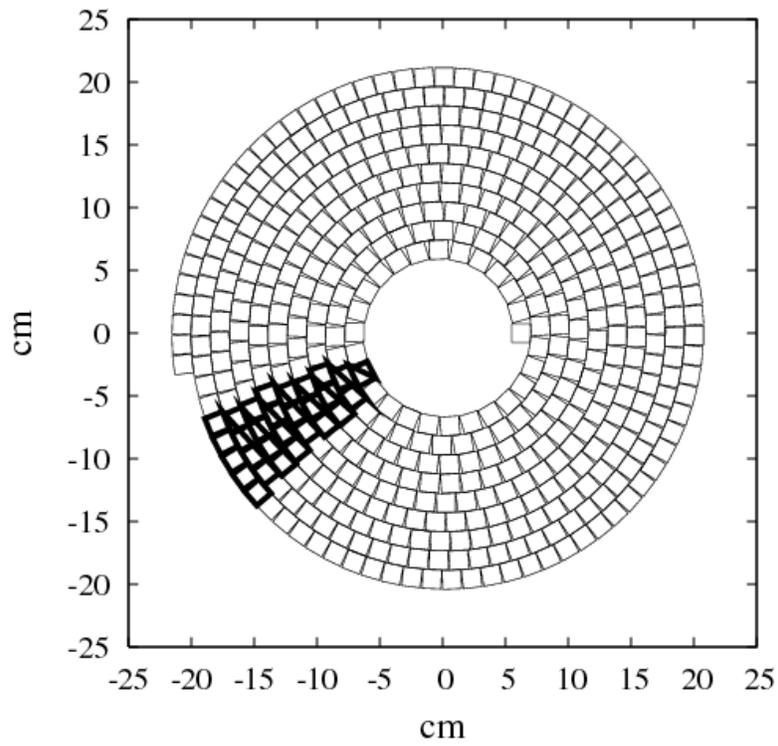}
\caption{Scheme of a lens of 210 cm focal length made of Cu(111) mosaic crystal 
tiles disposed according to an Archimedes' spiral. Also shown, in bold-faced, is 
the DM  we are developing to establish the crystal 
assembling technique.}
\label{f:DM}
\end{center}
\end{figure}

\begin{figure}
\begin{center}
\includegraphics[angle=0,width=0.6\textwidth]{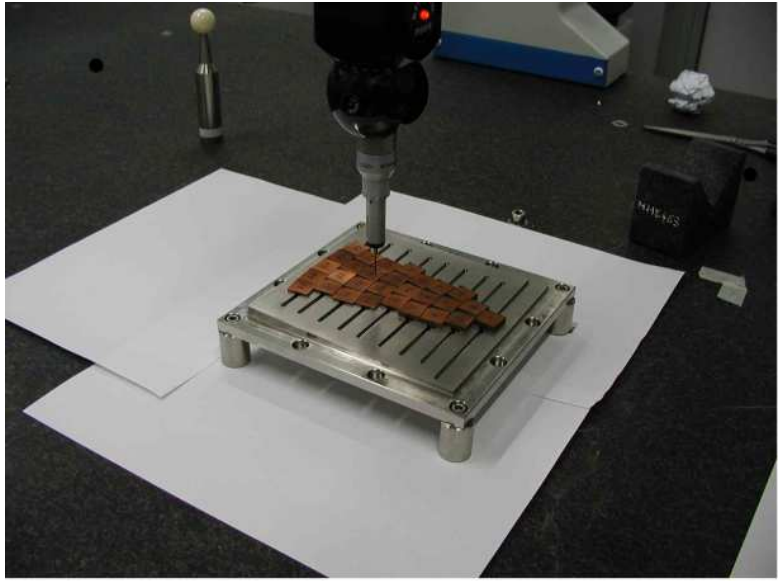}
\caption{Picture of the first DM mock-up during the testing phase of the positioning
accuracy with a profilometer.}
\label{f:mock-up}
\end{center}
\end{figure}

We are now developing a Demonstration
Model (DM) to establish the best crystal assembling technique of the lens. 
The DM  is a sector of the lens and is shown in Fig.~\ref{f:DM}. It is
composed of 30 mosaic crystal tiles with 3 arcmin spread and 15$\times$15 mm$^2$ 
front surface. We have already built some mock-up models, made of polycrystalline 
copper, and we are now starting to build a model made of true mosaic
crystals. The picture of the  
first mock-up assembly during the testing of the crystal positioning accuracy
is shown in Fig.~\ref{f:mock-up}. After the DM, we plan to build a Prototype
Model (PM), like the entire lens shown in Fig.~\ref{f:DM}.
The realization of the PM will be a good benchmark about our capability 
of scaling the assembling technology to larger lenses. Both DM and PM  
will be tested at the  Ferrara X--ray facility 
which is now being extended for this project \cite{Loffredo06}.
%
% Figure 10
%

\section{Study of possible lens configurations for the GRI mission}
\label{s:examples}

On the basis of the feasibility study above summarized and the test results 
of the mosaic crystal samples of Cu(111), we are now investigating different 
configurations of Laue lenses that meet the requirements 
of the GRI mission now under study. For a description of the science goals 
of this mission and its sensitivity requirements see Knoedlseder \cite{Jurgen06}.
Shortly, the main requirement of this mission is an 
unprecedented sensitivity for the study of  the continuum emission from about 
100 keV to 600 keV, and for the detection of both 511 keV annihilation lines and
nuclear lines in the range from 800 to 900 keV. For an observation time of $T = 10^5$~s, the
spectrum determination sensitivity should be of the order or better than 
$10^{-7}$~photons/(cm$^2$~s~keV) at 300 keV, while
the sensitivity to nuclear lines should be of the order of  
$10^{-6}$~photons/(cm$^2$~s) or better. The energy passband 
of the mission is required to be extended to lower energies with an X--ray monitor 
capable of determining the X--ray spectra of the gamma--ray target sources 
down to $\le$20 keV. The baseline monitor assumed is a coded mask telescope, with the mask to be 
located in the hole left free by the inner lens, i.e., the lens devoted to the nuclear lines 
(Nuclear Line Lens, NLL).

We have assumed lens sizes compatible with the fairing of the Russian
launcher Soyuz (maximum diameter of 380 cm and height of 925 cm). 
The best focal length that meets the continuum and nuclear line sensitivity 
requirements, is resulted to be of 75 m. A lower focal length would privilege 
the continuum sensitivity, while a higher focal length would privilege the 
nuclear line sensitivity.

As lens materials we have assumed only Cu(111) and Ge(111), which at the
moment are available with a mosaic structure. Other candidate mosaic crystals,
like those shown in Fig.~\ref{f:materials}, could become available in 
the next future.

In spite that Cu(111)  shows higher reflectivity 
than Ge (111) (see Fig.~\ref{f:materials}), due to  its low lattice spacing 
(2.087~$\AA$), a lens based on this material with 75 m focal length has an energy 
threshold (236 keV) that does not meet
the GRI requirements. These requirements can be met by a Ge(111) lens (energy 
threshold of 150 keV), which however is less efficient. In the first two columns of 
Table~\ref{t:lenses} we show the main features of two lenses suitable to study the continuum
emission (continuum lenses), one based on  Ge (111) (Low Energy Lens, LEL)
and the other based on Cu (111) (High Energy Lens, HEL).
It is apparent the better effective area (@ 300 keV) of the Cu lens with 
respect to the Ge lens, which however allows a lower energy passband. 

\begin{figure}
\begin{center}
\includegraphics[angle=-90,width=0.6\textwidth]{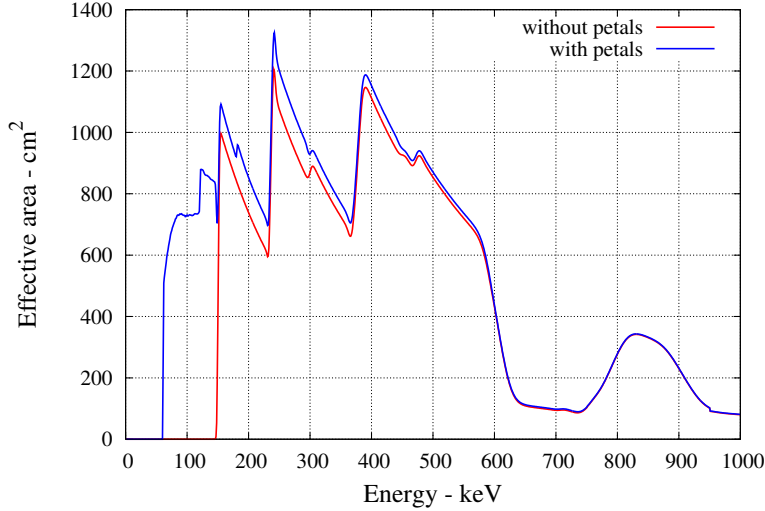}
\caption{Effective area of a broad energy passband lens (BEL) made of
Ge(111) and Cu(111) to cover the 150 to 600 keV plus a nuclear line lens
(NLL) to cover the 800 to 900 keV energy band (see features in Table~\ref{t:lenses}). 
Also shown is the effective area of the same lens, with energy passband extended down
to 60 by means of the addition of a 4 petal-like lens (see features in 
Table~\ref{t:lenses}).}
\label{f:lenses}
\end{center}
\end{figure}

To get a Broad Energy Lens (BEL) with a passband from 150 to 600 keV, we have
assumed a lens partly covered with Ge(111) and partly with Cu(111),
with an unavoidable loss of effective area at low energies with respect to the HEL/Cu
(see Table~\ref{t:lenses}).
Figure~\ref{f:lenses} shows the effective area of the BEL plus that of the NLL, which is 
designed for the 800--900 keV passband. The NLL features are 
shown in Table~\ref{t:lenses}. For the latter lens, unlike a mosaic spread of 1 arcmin,
we have assumed 40 arcsec, to decrease the half power radius of the focused photon 
spot (see Table~\ref{t:lenses}). 

We have also investigated the possibility of extending the energy passband of the 
BEL from 150 keV down to 60 keV, through the addition of a petal-like lens, 
to be deployed in orbit. 
In Table~\ref{t:lenses} we show the main features this lens, which is made of 4 petals of 
100 cm width. The effective  area of this lens and its impact to higher energies
is shown in Fig.~\ref{f:lenses}. Such additional lens could become mandatory if
the X--ray monitor does not meet the sensitivity requirements. 

%
% Figure 11
%

\begin{figure}
\begin{center}
\includegraphics[angle=-90,width=0.6\textwidth]{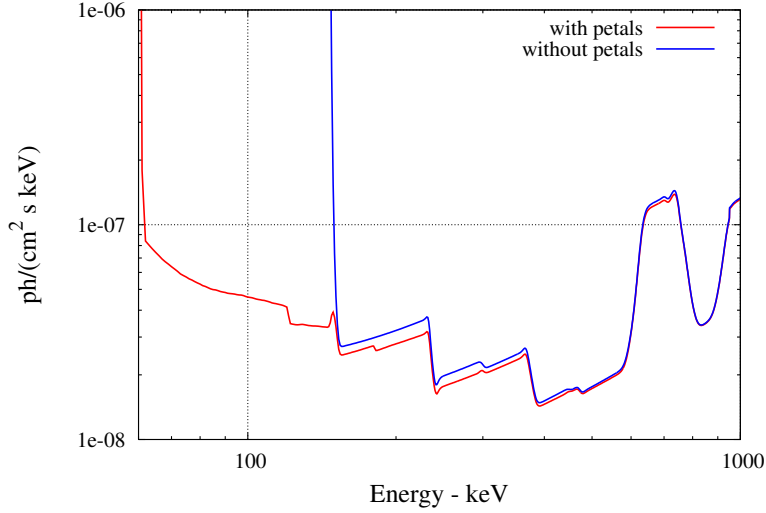}
\caption{Expected ensitivity at a $3\sigma$ confidence level of the BEL
 plus NLL lenses with (lower line) or without (upper line) the addition of
a petal-like lens with 4 petals of 100 cm width (see details in the text and in the
 Table~\ref{t:lenses}).}
\label{f:sensitivity}
\end{center}
\end{figure}

The expected 3$\sigma$ sensitivity to the continuum emission
of the BEL plus NLL with or without petals
is shown in Fig.~\ref{f:sensitivity}. We have assumed an exposure
time of $10^6$~s, and a ratio $R = \sqrt{B/\eta_d} = 7\times 10^{-3}$. To get
this figure, it is crucial to have a low detector background and 
a high detection efficiency. The expected
sensitivity to the 511 keV (847 keV) line is given by $I_{511} = 1.9\times 
10^{-7}$~photons~cm$^{-2}$~s$^{-1}$ ($I_{847} = 1.4\times 10^{-6}$~photons~cm$^{-2}$~s$^{-1}$),
for an assumed ratio $R = 1\times 10^{-2}$ and line width of 3 keV (25 keV).

%
% Figure 12
%

\begin{table}
\caption{Main features of the different Laue lens configurations}
\label{t:lenses}    
\begin{tabular}{llllll}
\hline\noalign{\smallskip}
Parameter & LEL/Ge & HEL/Cu & NLL/Cu  & BEL/Ge+Cu & Petals/Ge \\	
\noalign{\smallskip}\hline \hline \noalign{\smallskip} \\
Focal length (m) &  75  &  75 &  75 & 75 &  75 \\ 
Nominal passband (keV) & 150--385  &  236--600 & 800--900 & 150--600 & 60--150 \\
Inner radius (cm) & 74   & 74  &  50  & 74 & 193 \\
Outer radius (cm) & 190  & 190 &  57  & 190  & 475 \\
Crystal material &  Ge(111) & Cu(111) &  Cu(111) & Ge(111) + Cu(111) &  Ge(111) \\
Mosaic spread (arcmin) & 1  &   1     & 0.67     & 1                 &  1  \\
Tile cross section (mm$^2$) & $10 \times 10$  & $10 \times 10$  & $10 \times 10$ &
    $10 \times 10$ &  $10 \times 10$  \\
Tile thickness (mm)    &  optimized  & optimized & 9  & optimized  & optimized  \\
No. of tiles & 89197 &  89197  & 2242  & 31865 (Ge), 57191 (Cu) & 108952 \\ 
Lens total surface (m$^2$) &  9.62  & 9.62 & 0.235  & 9.62  & 11.3 \\ 
Filling factor & 0.93  & 0.93 & 0.95  &  0.95  &  0.96  \\     
Crystal weight (kg)  & 115 & 175 & 18 & 200  & 58 \\
Effective area (cm$^2$) @ 300 keV &  862   & 1652  &  -  &  870  &  60   \\ 
Effective area (cm$^2$) @ 511 keV &  160   &  932  &  -  &  854  &  17   \\
Half power radius(mm) &  14  &   14   &  9   & 15   &   15  \\
\noalign{\smallskip}
\hline
\end{tabular}
\end{table}

\section{Conclusions}
The development status of a Laue lens project has been outlined.
After a design phase of a broad band Laue lens based on mosaic crystals,
we have passed to an experimental phase with the test of
mosaic crystals of Cu(111). All the results obtained have been very promising. 
We are now developing a first lens prototype. The Monte Carlo simulations show
interesting expectations of the Laue lenses designed, with an expected angular
resolution of about 2 armin and a field of view that depends on the focal 
detector radius and instrument sensitivity. We have insestigated possible lens 
configurations for the GRI mission, finding that the best compromize for getting 
an unprecedented sensitivity for the continuum spectrum determination in the 
150--600 keV and for the emission line detection (511 keV and lines in the band from 
800 to 900 keV) is a set of two lenses of 75 m focal length, one made of Ge(111) 
and Cu(111) that covers the band from
150--600 keV, and the other made of Cu(111) with 800--900 keV passband.
The lens passband can be extended to 60 keV with the addition of a petal-like lens. This
lens is also compatible with a Soyuz launcher fairing, but needs to be deployed in orbit. This
deployement  capability is estimated to be feasible.

\acknowledgments     %>>>> equivalent to \section*{ACKNOWLEDGMENTS}       
 
We acknowledge the support by the Italian Space Agency and Italian Institute of
Astrophysics (INAF). This research was also possible thanks to received Descartes
Prize 2002 of the European Committee.

%%%%%%%%%%%%%%%%%%%%%%%%%%%%%%%%%%%%%%%%%%%%%%%%%%%%%%%%%%%%%
%%%%% References %%%%%

\end{document}